\documentclass[10pt]{iopart}
\pdfoutput=1
\usepackage{graphicx}
\usepackage{appendix}

\usepackage{iopams}
\usepackage{comment}
\usepackage{xcolor}

\begin{document}

\title{Criticality in Tissue Homeostasis: Models and Experiments}
\author{Somsubhra Ghosh}
\address{Department of Physics, University College of Science and Technology, Kolkata-700009, India}
\author{Indrani Bose}
\address{Department of Physics, Bose Institute, Kolkata-700009, India}
\ead{indrani@jcbose.ac.in}
\vspace{10pt}

\begin{abstract}
There is considerable theoretical and experimental support to the proposal that tissue homeostasis in the adult skin can be represented as a critical branching process. The homeostatic condition requires that the proliferation rate of the progenitor (P) cells (capable of cell division) is counterbalanced by the loss rate due to the differentiation of a P cell into differentiated (D) cells, so that the total number of P cells remains constant. We consider the two-branch and three-branch models of tissue homeostasis to establish homeostasis as a critical phenomenon. It is first shown that some critical branching process theorems correctly predict experimental observations. A number of temporal signatures of the approach to criticality are investigated based on simulation and analytical results. The analogy between a critical branching process and mean-field percolation and sandpile models is invoked to show that the size and lifetime distributions of the populations of P cells have power-law forms. The associated critical exponents have the same magnitudes as in the cases of the mean-field lattice statistical models. The results indicate that tissue homeostasis provides experimental opportunities for testing critical phenomena.
\end{abstract}
\vspace{2 pt}

%

\section{Introduction}
A characteristic feature of adult mammalian tissues is that of homeostasis implying steady state conditions \cite{klein2011universal,jones2008opinionepidermal}. During the developmental stage of an organism, a tissue increases in size as a function of time due to a proliferation in the number of cells, constituting the tissue, through repeated rounds of cell division. In the adult tissue, the number of cells capable of undergoing cell division, designated as progenitor (P) cells, remains constant giving rise to an unchanging tissue size. A pioneering experiment on the fate of cells in the tail epidermis of mice provides the basis for a simple model of homeostasis \cite{clayton2007single,klein2007kinetics}. The epidermis is the outermost of the three layers of tissues that make up the skin. It consists of a basal layer and a few supra-basal layers. Two types of cells are present in the basal layer: the P cells and the differentiated (D) cells, with only the P cells undergoing cell division. The P cells undergo cell division with three possible outcomes: PP (both daughter cells are P cells), PD (one daughter is a P cell while the other is a D cell) and DD (both the daughters are D cells). The probabilities for these three outcomes are $a$,$b$ and $c$ respectively with $a+b+c=1$. The D cells migrate from the basal to the supra-basal layers and are finally shed from the surface of the skin. The condition for homeostasis in the basal layer is $a=c$, i.e., the proliferation rate of the P cells is counterbalanced by the loss rate of P cells due to differentiation so that the total number of P cells remains constant. We designate the model of homeostasis as the three-branch model (figure 1(a)). A simpler version of the model with the same qualitative behaviour is the two-branch model  (figure 1(b)) in which the probability of asymmetric cell division, $b=0$. These models are similar to the models studied earlier to investigate the dynamics of early tumour growth \cite{remy2016near}.\\

\noindent The colony of P cells that grows from a single progenitor defines a branching process (figure 2) \cite{roshan2014exact,rue2015cell}. The theory of branching processes has largely been developed by mathematicians with several powerful theorems and rigorous results proved and derived over the years \cite{harris2002theory,zbMATH03410334,kimmelbranching}. The applications of the theory are wide-ranging, from cosmic ray showers and nuclear chain reactions to the growth of reproducing populations. Examples of the latter include animals, plants, bacteria, royal families etc. In fact, the branching process model was originally conceived to determine the number of generations in which the British royal family name, with inheritance passing from the father to the son, would possibly become extinct. In the branching process model, there are three distinct dynamical regimes: subcritical $(a<c)$, critical $(a=c)$ and supercritical $(a>c)$.  In the subcritical case, the population of P cells becomes extinct, i.e., no P cells are left in the course of time. The probability $q$ for eventual population extinction is given by $q=1$ in this case. In the supercritical case, the probability  $q$ is non-zero but less than one, opening up the possibility of indefinite growth of the population. At the critical point, $a=c$, the time evolution of the population has features distinct from those of the subcritical and supercritical regimes. The variance of the distribution of the population size of a critical branching process grows linearly as a function of time and the large fluctuations are responsible for population extinction with probability $q=1$ in the limit of large times. The state in which the number of P cells is zero is the so-called absorbing state from which revival of the population is not possible. In section 2 of the paper, we describe the Galton-Watson (GW) model of a branching process \cite{harris2002theory,zbMATH03410334,kimmelbranching} and state a few theorems and results relevant for our study of tissue homeostasis. We point out the utility of the theorems in providing an understanding of experimentally observed phenomena on tissue homeostasis. Using one of the theorems, we show that the cumulative distribution function (CDF) of the colony of P cells is that of a gamma distribution in the case of the critical branching process $a=c$.\\

\begin{figure}
\begin{center}
\begin{minipage}[c]{0.99\linewidth}
\includegraphics[width=13.2cm]{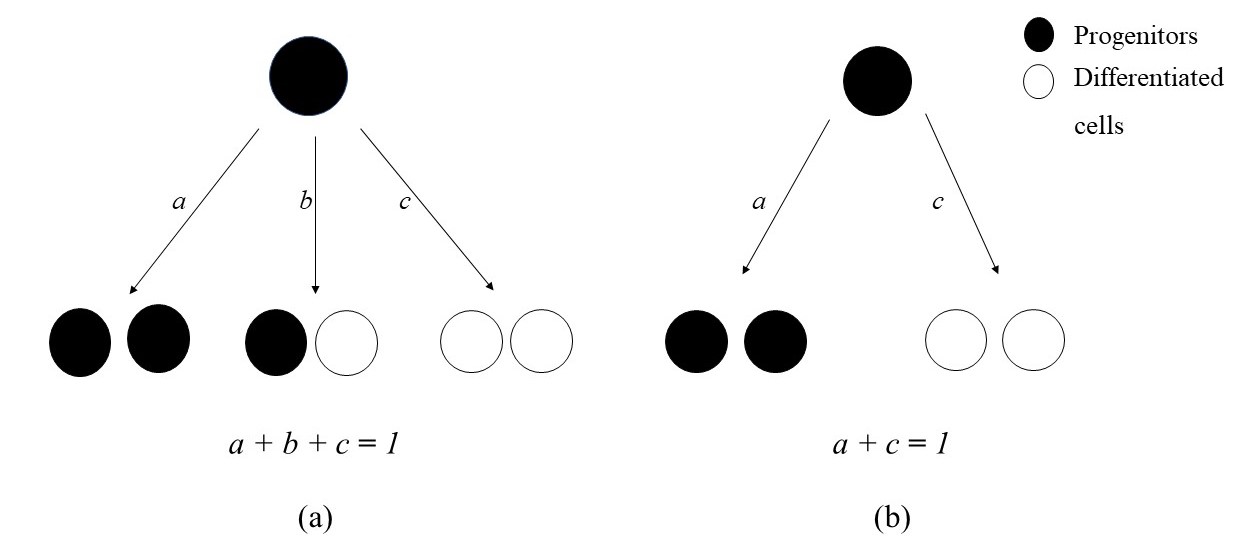}
\end{minipage}

\caption{Schematic representation of a (a) three-branch and a (b) two-branch process. The parameter associated with each branch gives the probability of cell division via that branch.}
\end{center}
\label{br1}
\end{figure}
\noindent Critical point transitions in the equilibrium and non-equilibrium are characterized by a number of features, collectively known as critical phenomena, exhibited close to criticality and at the critical point itself \cite{goldenfeld2018lectures,sornette2006critical}. In section 3, we present Monte Carlo (MC) simulation results for some quantitative signatures of the approach to the critical point. These include the variation of the mean time to extinction and the mean time to reach a threshold population size as a function of the ratio of parameters $c/a$,  with $c/a=1$ at the critical point. We further study the distributions of the time to extinction and the time to reach the threshold population size, as well as the variances of the distributions as a function of $a$. We show that the simulation results, obtained in the case of the discrete-time (DT) GW process, are in qualitative agreement with the analytical results derived by treating the branching process as continuous-time (CT). In the case of the two-branch model, a quantitative comparison, made possible due to the property of embeddability, is also carried out. Statistical physics models like the sandpile model of self-organised criticality (SOC) and the percolation model exhibit critical phenomena which, in the mean-field limit (fluctuations ignored) can be described in terms of a critical branching process \cite{alstrom1988mean,zapperi1995self,gros2010complex}. Keeping this equivalence in mind, tissue homeostasis, an example of a critical branching process, provides experimental opportunities for testing critical phenomena predictions. In section 4, we make use of the generating function for the total number of P cells (branching events) produced to illustrate critical phenomena similar to those exhibited by the sandpile and percolation models. Section 5 contains a summary of the main results obtained in the paper and some concluding remarks.

\begin{figure}
\begin{center}
\begin{minipage}[c]{0.99\linewidth}
\includegraphics[width=13.2cm]{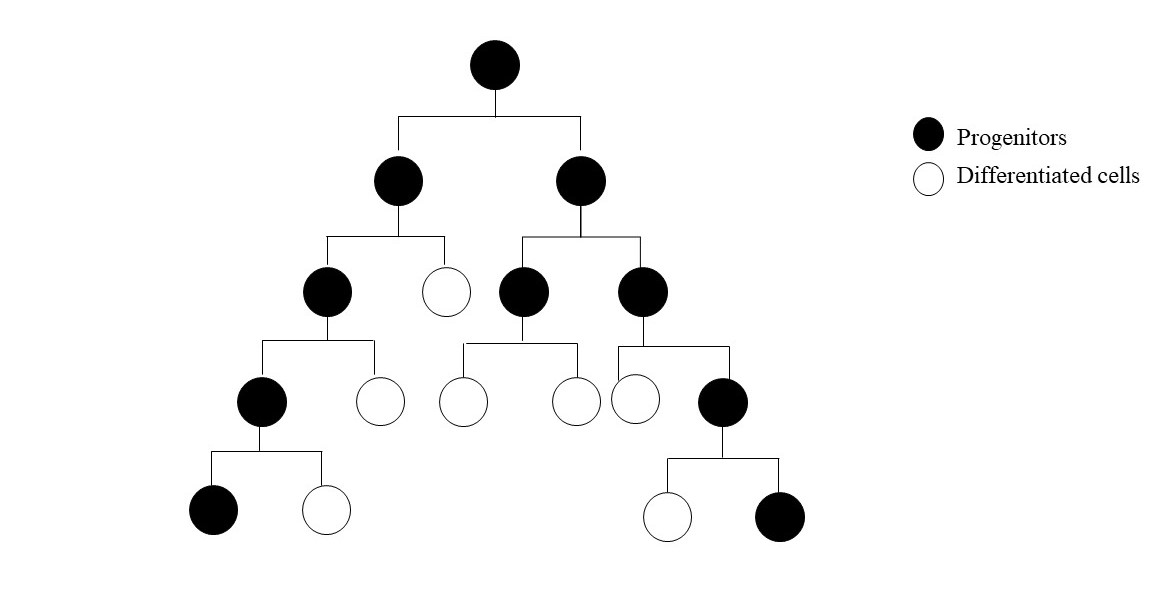}
\end{minipage}

\caption{Lineage of a single progenitor (P) cell represented as a branching process. The different branches progress independent of each other. The type of division (symmetric or asymmetric) at each branching point is a probabilistic process.}
\end{center}
\label{br2}
\end{figure}
\section{Branching Process Theorems and Experimental Validation}
We first provide a brief description of the GW branching process \cite{harris2002theory,zbMATH03410334}. One assumes that a single individual is present at time $t=0$. The individual lives for one unit of time (time advances in discrete steps). At time $t=1$, the individual produces a family of offspring and immediately dies. The number of offspring is a random variable and defines the family size $Y$. The family size/offspring probability distribution is given by

\begin{equation}
    P(Y=k)=p_k, \hspace{20 pt} k=0,1,2,...
    \label{famprob}
\end{equation}
\noindent The general rule is that each individual, existing at time, $t-1, (t=1,2,3,...)$ lives for one unit of time, produces its own family of offspring at time $t$ and immediately dies. One assumes that all individuals reproduce independently of each other and the family sizes of different individuals are independent random variables with the probability distribution given in equation (\ref{famprob}).\\

\noindent We next introduce the concept of the probability generating function (PGF), useful to deal with probability distributions and their moments. Let $X$ be a random variable taking non-negative integral values \{$0,1,2,...$\} with a specific probability distribution. The probability generating function (PGF) of $X$ is given by

\begin{equation}
    G_X (s)=E(s^X)=\sum_{k=0}^{\infty}s^k P(X=k)
\end{equation}
\noindent where $E(..)$ denotes the expectation value or average. Some well-known properties of the PGF are:

\begin{eqnarray}
G_X (1)=1,\nonumber\\
E(X)=G'_X (1), \label{gf_prop}\\ 
E[X(X-1)]=G''_X (1) \nonumber
\end{eqnarray}
\noindent where the prime symbol denotes differentiation with respect to $s$. In the case of the three-branch model of tissue homeostasis, the PGF of the family size $Y$  of each progenitor is

\begin{equation}
    G(s)=as^2+bs+c
    \label{pgf}
\end{equation}
\noindent with $G(1)=a+b+c=1$. From equation (\ref{gf_prop}), the mean family size distribution, i.e., the average number of P cells produced per progenitor, $m$ is given by

\begin{equation}
    m=G'(1)=2a+b=a+1-c
\end{equation}
\noindent The condition for homeostasis is $a=c$ (critical branching process) yielding $m=1$. Also, for $a>c$ (supercritical branching), $m$ is $>1$ and for $a<c$ (subcritical branching), $m$ is $< 1$. Let $\sigma^2$ be the variance of the family size distribution and $Z_t$ be the size of the population, i.e., the total number of individuals at time $t$. One can then derive the following results for the mean and variance of $Z_t$:

\begin{eqnarray}
E(Z_t)=m^t \nonumber \label{moment} \\ 
Var(Z_t)=\sigma^2t \hspace{70 pt} if \hspace{5 pt}m=1 \\
Var(Z_t)=\sigma^2m^{t-1}\frac{1-m^t}{1-m} \hspace{20 pt} otherwise \nonumber
\end{eqnarray}
\noindent An issue of interest in branching process dynamics is that of the extinction of the population. The population of P cells becomes extinct at time $t$ if $Z_t=0$ but the size of the population is non-zero at earlier time points. Once extinction occurs, the population size continues to remain zero at all future times. We define $q$ to be the probability of population extinction. A branching process theorem states \cite{harris2002theory} that the probability $q$ is the smallest non-negative solution of the equation:

\begin{equation}
    G(s)=s
    \label{ext}
\end{equation}
\noindent where G(s) is the PGF of the family size distribution, which for the three-branch model is given by equation (\ref{pgf}). The solutions of equation (\ref{ext}) turn out to be
\begin{eqnarray}
    q=\frac{c}{a} \hspace{20 pt} for \hspace{5 pt}c<a\hspace{2 pt}(m>1) \nonumber\\
    q=1 \hspace{20 pt} for \hspace{5 pt}c>a\hspace{2 pt}(m<1) \label{eqn8}\\
    q=1 \hspace{20 pt} for \hspace{5 pt}c=a\hspace{2 pt}(m=1) \nonumber
\end{eqnarray}
The results show that population extinction is certain in the subcritical and critical cases whereas it has a finite probability ($q<1$) in the supercritical case. The results for the subcritical and supercritical processes can be understood from the expression of the average population size, $E(Z_t)$, at time $t$ (equation \ref{moment}). When $m$ is $<1$, the term $m^t$ $\rightarrow$ $0$ as $t$ becomes large. When $m$ is $>1$, there is a finite probability for indefinite growth of population as time progresses. In the critical branching case $(m=1)$, large fluctuations (variance grows linearly as a function of time) are responsible for the eventual extinction of the population. Thus, irrespective of the nature of the branching process, the sequence of population sizes, $\{Z_t\}$,  either goes to zero (extinction) or to $\infty$ (explosion) in the limit of large times, i.e., 
\begin{equation}
    \lim_{t\to\infty} P(Z_t = k)=0, \hspace{20 pt} k=1,2,3,...
    \label{eqn9}
\end{equation}
where $k$ has a finite, non-zero value. The fate of the population in the limit of large time is thus between extinction and explosion so that
\begin{equation}
    P(Z_t\to0) + P(Z_t\to \infty)=1
    \label{eqn10}
\end{equation}
with the respective probabilities of the two processes being $q$ and $1-q$.\\

\noindent We next state three theorems \cite{harris2002theory,kimmelbranching} for the critical branching process $a=c$ which we show to be consistent with the experimental results on tissue homeostasis.\\
Theorem 1: If $m=1$ and $G'''(1)<\infty$, then in the limit of large $t$:
\begin{equation}
    P(Z_t>0)\approx\frac{2}{tG''(1)}
    \label{Th1}
\end{equation}

\noindent Since, for the critical branching process, $E(Z_t)=1=E(Z_t|Z_t\neq 0)P(Z_t\neq 0)$, we can utilize Theorem 1 to write\\
Theorem 2: 
\begin{equation}
    E(Z_t|Z_t\neq 0)\approx\frac{tG''(1)}{2}
    \label{mean}
\end{equation}
Theorem 3: If $m=1$ and $G'''(1)<\infty$, then in the limit of large $t$:
\begin{equation}
P(\frac{Z_t}{t}>u|Z_t>0)\approx \exp(-\frac{2u}{G''(1)}), \hspace{20 pt} u\geq0
\label{Th3}
\end{equation}
For the two-branch and three-branch models of tissue homeostasis, $G''(1)=2a=\sigma^2$, the variance of the family size distribution.\\

\noindent Experimental observations by Clayton et al. \cite{clayton2007single,klein2007kinetics} on tissue homeostasis in adult tail epidermis of mice are in agreement with the contents of Theorems 1-3, pertaining to a critical branching process. In the experiment, starting with a single labelled cell and using the techniques of genetic lineage tracing, the time evolution of the progeny population could be tracked  with single cell resolution. Let $P_n(t)$ be the probability that the number of P cells present at time $t$ is $n$. The Master Equation (ME) for the probability distribution is amenable to exact, analytic solution  for the two-branch model, $b=0$, and with $a=c=1/2$ .  With the analytic expression for $P_n(t)$ known in this case, the average number of P cells at time $t$ is found to be $\langle n\rangle=\sum_{n\geq1}nP_n(t)=1$. At this point, we note that the critical branching process theory yields the result $\langle n\rangle=E(Z_t)=m^t=1$ (equation (\ref{moment}) with $m=1$) under the more general conditions, $b\neq0$ and $a=c$.  According to Theorems 1 and 2, the survival probability of a progeny population is given by $\sim \frac{1}{at}$, whereas the average size of persisting clones increases as $\sim at$ at large times. These results are true for both the two and three-branch models and are in accordance with the experimental observations by Clayton et al. \cite{clayton2007single}. In the case of the two-branch model, the ME approach and the critical branching process theorems yield the same results for $at\gg1$. The mathematical results lead to the understanding that tissue homeostasis, in terms of the average number of P cells remaining constant, is achieved due to the compensation of a continual extinction of clonal populations by the steady growth of persisting clonal populations.\\

\noindent A noteworthy feature of the clone size distribution, measured experimentally, in the mice epidermis, is the collapse of the data onto a single scaling curve in the limit of large time. The scaling form of the distribution on persisting clones is given by \cite{klein2011universal}
\begin{equation}
    P_{n}^{pers}=\frac{\tau}{t}f(\frac{n\tau}{t})
\end{equation}
with $f(x)=e^{-x}$. The scaling form implies that the probability of finding a clone size in between $\frac{n}{2}$ and $n$ cells at time $t$ is the same as that of finding a clone size in between $n$ and $2n$ cells at time $2t$. The scaling form is consistent with that provided by Theorem 3 (equation (\ref{Th3})) with the parameter $\tau=1/a$.In the experiment, the lineage tracing technique implemented through the labelling of cells does not distinguish between the P and D cells, with $n$ indicating the total number of cells. We show at the end of section 4 that this does not change the basic results obtained by treating $n$ as the number of P cells.\\

\noindent The earliest stochastic model of cell proliferation and differentiation was proposed by Till et al.\cite{till1964stochastic}, based on their pioneering experiment involving spleen colony assay in mice. They noticed that the colonies have a heterogeneous distribution of the number of colony-forming cells (designated as colony forming units or CFUs) with only a few colonies containing a large number of CFUs. Till et al. analysed the experimental data in terms of a model similar to the two-branch model of cell proliferation and differentiation (figure \ref{br1}(b)) studied in this paper. The experimental data on the cumulative distribution function (CDF) of the CFUs per colony could be fitted well by that of a gamma distribution, with the distribution having the same mean and variance as the experimental data. The data also agreed closely with the Monte Carlo (MC) simulation results of the stochastic birth-death model. The MC calculations assumed fixed birth and death probabilities and a fixed generation time, as in the usual branching process model. Based on the literature available at the time \cite{harris1959kinetics}, Till et al. had conjectured that the stochastic birth-death model generates a CDF, well-approximated by that of the gamma distribution, independent of the distribution of generation times.  We now show, invoking Theorem 3, that the CDF in the case of a critical branching process is indeed that of the gamma distribution. From  Theorem 3, one obtains, with $G''(1)=2a$,
\begin{equation}
    P(Z_t>u|Z_0 = 1, Z_t > 0) \sim \exp(-\frac{u}{at})    \label{clayton}
\end{equation}
with $u\geq0$. Thus, the CDF of the distribution of P cells at the $t$-th generation is
\begin{equation}
    P(Z_t\leq u|Z_t\neq 0) = 1 - \exp(-\frac{u}{at})
    \label{cdf}
\end{equation}
The CDF of a gamma distribution with shape parameter $k$ and mean $k\theta$ has the form
\begin{equation}
    F(x;k,\theta)=\frac{1}{\Gamma(k)}\int_{0}^{x/\theta} y^{k-1}e^{-y} dy
    \label{cdfgamma}
\end{equation}
where $\Gamma (k)$ is the gamma function. A comparison of equations (\ref{cdf}) and (\ref{cdfgamma}) shows that the CDF in a critical branching process (equation (\ref{cdf})) is that of a gamma distribution with $k=1$ and the mean $k\theta=at$, the mean of the probability distribution of surviving clones of P cells (equation (\ref{mean}) with  $G''(1)=2a$). Figures 3(a) and 3(b) show the MC simulation CDF data (represented by dots) in the cases of the two-branch ($b=0,a=c=0.5$) and three-branch ($b=0.4,a=c=0.3$) models respectively through 20 generations and 1000 simulation runs. The solid lines correspond to the CDF of the gamma distribution with form as in equation (\ref{cdf}).
\begin{figure}
\begin{center}
\begin{minipage}[c]{0.99\linewidth}
\includegraphics[width=13.2cm]{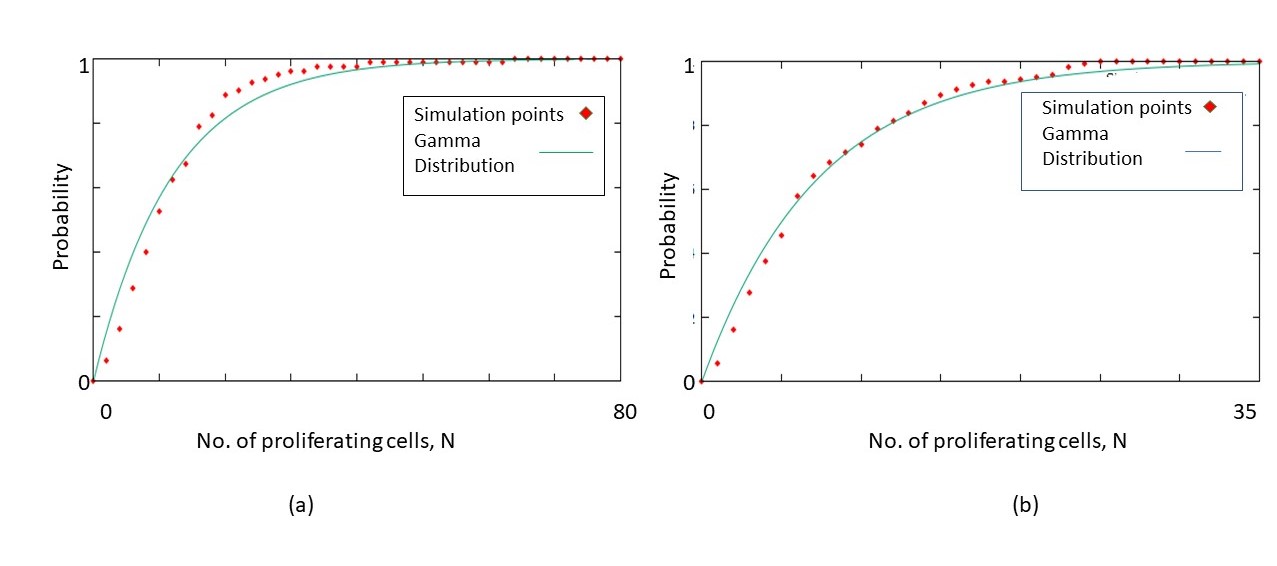}
\end{minipage}
\caption{Simulation results (red diamonds) for cumulative probability distribution of P cells at the $20^{th}$ generation for (a) a=0.5,b=0,c=0.5 and (b) a=0.3,b=0.4,c=0.3. The solid line in each case depicts the fitted gamma distribution.}
\end{center}
\label{CDFMonteCarloGamma}
\end{figure}
\section{Temporal Signatures of Approach to Criticality}
\noindent The critical branching process describing tissue homeostasis satisfies the condition $a=c$ for the branching probabilities with the mean number of offspring per individual $m=1$. Regardless of the value of $m$, any state with finite population size $k\neq0$ is transient \cite{harris2002theory,zbMATH03410334} (equation (\ref{eqn9})). In the large time limit, the fate of a population of P cells is either extinction or explosion (equation (\ref{eqn10})). We now show that the approach to the critical point $\frac{c}{a}=1$ carries distinctive temporal signatures in terms of quantities like the mean extinction time, $T_{ex}$, and the mean time, $T_{th}$, to reach a threshold population size $N_{max}$. We use the same MC simulation procedure as discussed in Ref. \cite{remy2016near} for our investigation. A brief description of the procedure is as follows. For specific values of the parameters $a$ and $c$, a MC simulation run yields time series data for the population size (number of P cells) growing from $Z_0=1$ to $N_{max}$ or to extinction. A fraction of the total number of simulation runs $S_{tot}$ results in extinction for which the mean extinction time, $T_{ex}$, is calculated. For the rest of the runs, the population attains the threshold size $N_{max}$ and one calculates the mean time, $T_{th}$, to reach the threshold size. In our simulation, we set the values $N_{max}=10000$ and $S_{tot}=10000$.\\

\begin{figure}
\begin{center}
\begin{minipage}[c]{0.99\linewidth}
\includegraphics[width=13.2cm]{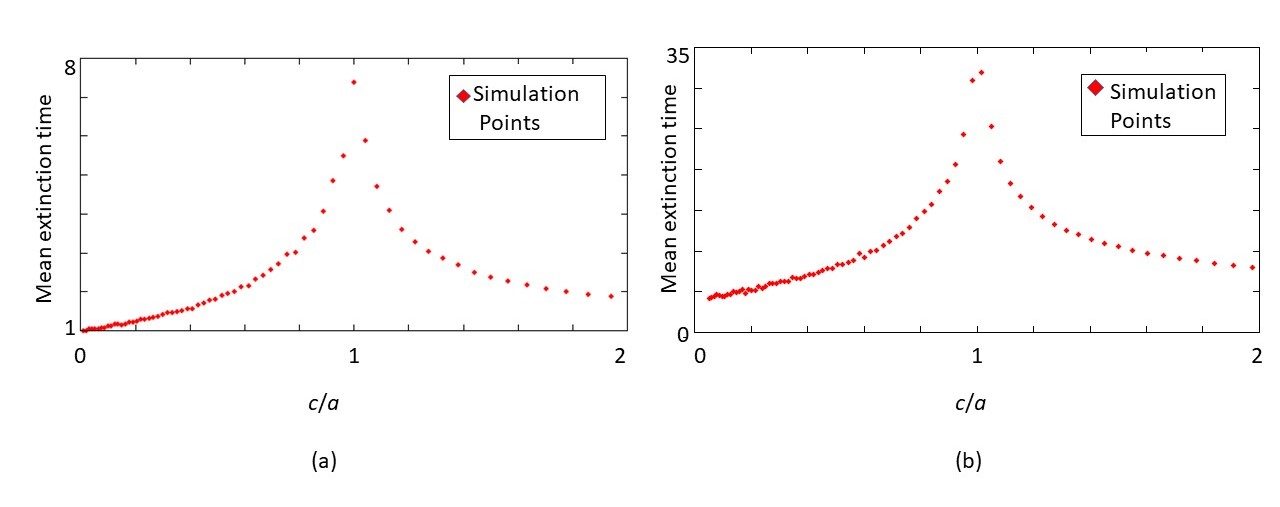}
\end{minipage}
\caption{Variation of mean extinction time $T_{ex}$ versus $\frac{c}{a}$ for (a) $b=0$ and (b) $b=0.75$. It is seen that $T_{ex}$ reaches a maximum at $\frac{c}{a}=1$ with the higher maximum occurring for $b=0.75$.}
\end{center}
\label{MeanTexSim}
\end{figure}
\noindent Figures 4(a) and 4(b) show the plots of $T_{ex}$ versus $\frac{c}{a}$  for $b=0$ and $b=0.75$ respectively. $T_{ex}$ reaches its maximum value when the branching process is critical, i.e, $\frac{c}{a}=1$. The maximum value of $T_{ex}$ increases with increase in the magnitude of $N_{max}$ and $T_{ex}$ diverges in the limit of $N_{max}\to\infty$. When $\frac{c}{a}\gg1$, the probability of generation of D cells is much greater than that of the P cells so that the mean extinction time for the population of P cells is small. As $\frac{c}{a}$ approaches $1$, $T_{ex}$ increases in magnitude as the probability of generation of the P cells becomes progressively closer to that of the D cells so that the average cell population size increases. At $a=c$, the mean extinction time becomes maximum tending to infinity as $N_{max}$ becomes infinitely large. In the case of $\frac{c}{a}\ll1$, the extinction probability decreases with most of the clusters exceeding the size limit $N_{max}$ with the few populations which go extinct, doing so within the first few generations, resulting in a small value of $T_{ex}$. In the case of the three-branch model, the non-zero value of $b$ has the effect of increasing the magnitude of the maximum of $T_{ex}$ (figure 4(b)) with $N_{max}$ having a finite value. An increase in $b$ implies an increase in the probability of the number of progenitor cells in the cell population remaining unchanged so that $T_{ex}$ reaches a higher maximum value.\\
\begin{figure}
\begin{center}
\begin{minipage}[c]{0.99\linewidth}
\includegraphics[width=13.2cm]{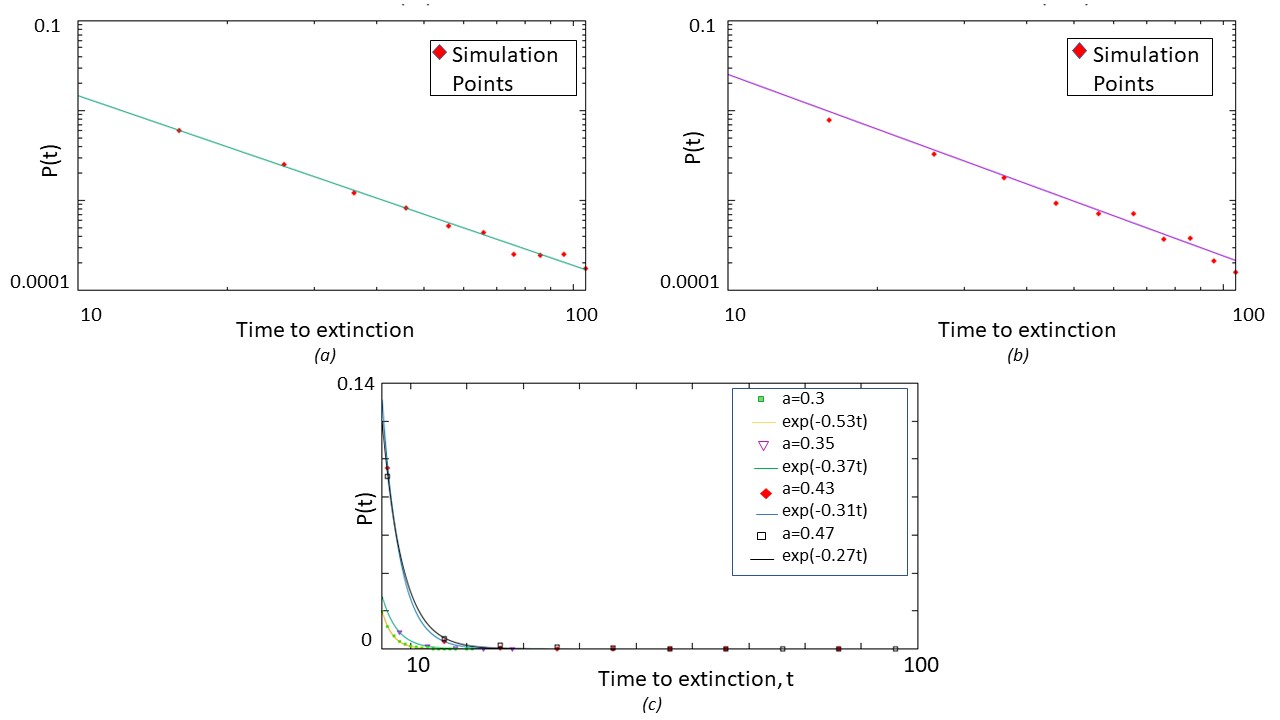}
\end{minipage}
\caption{Distribution of time to extinction for (a) $b=0, a=0.5$, (b) $b=0.25, a=0.375$ and (c) $b=0$,$a<0.5$. The first two cases represent this behaviour at criticality $(a=c)$ whereas the third case corresponds to the subcritical regime. In the critical case, the distribution has a power law form whereas it is exponential in the subcritical regime.} 
\end{center}
\label{DistExtTime}
\end{figure}

\noindent This behaviour of the mean extinction time can be understood, if one looks at the distribution of the time to extinction for both the critical and the off-critical cases (figures 5(a) $-$ 5(c)). For the critical case, the distribution shows a power law behaviour indicating an absence of a characteristic time scale in the system, with the mean diverging in a power law fashion. In the off-critical case, the distribution is exponential indicating the presence of a characteristic time scale, $t_c$. The mean extinction time in this case is given by 
\begin{eqnarray}
    T_{ex}\sim\sum_{t=1}^{\infty} t\exp(-\frac{t}{t_c}) \nonumber \\
    \hspace{14 pt} =\frac{\exp(-1/t_c)}{(1-\exp(-1/t_c))^2}
\end{eqnarray}
which is clearly finite for a finite $t_c$. It is expected that this characteristic time scale should decrease as we move away from criticality, so that the mean extinction time decreases. This behaviour is evident from figure 5(c).\\

\noindent The possibility of the population size reaching the threshold value $N_{max}$ is realized when $a$ is greater than $c$. Figures 6(a) and 6(b) show the variation of the mean time, $T_{th}$, for reaching the threshold size $N_{max}$ as a function of $\frac{c}{a}$ with $N_{max}=100,000$, $S_{tot}=10000$ and $b=0$ , $b=0.75$ respectively. The figures show that the mean 
\begin{figure}
\begin{center}
\begin{minipage}[c]{0.99\linewidth}
\includegraphics[width=13.2cm]{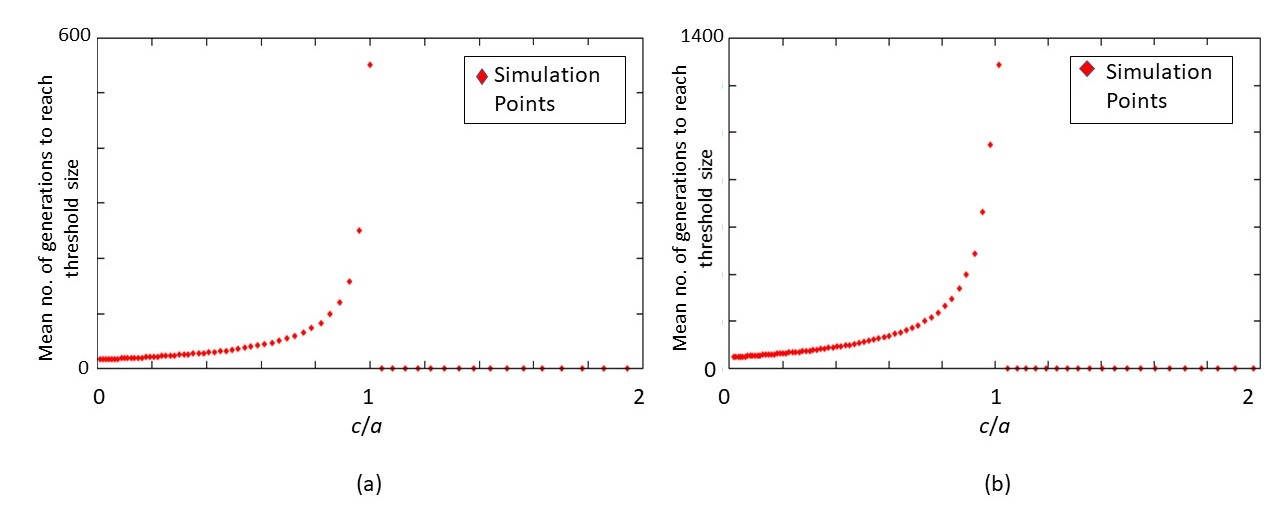}
\end{minipage}
\caption{Variation of mean time, $T_{th}$, taken by the population of P cells to reach the threshold size, $N_{max}=100,000$ versus $c/a$ for (a) $b=0$, (b) $b=0.75$. $T_{th}$ is seen to diverge as $c/a\to1$.} 
\end{center}
\label{MeanTthSim}
\end{figure}
\noindent time diverges as $\frac{c}{a}\to1$. The plots in figures 4 and 6 are similar to the ones in Ref. \cite{remy2016near} obtained in the case of evolving tumour cell populations. The plots obtained are based on simulation results in both the cases. An analytic expression for the distribution of times to reach a threshold size can be derived in the case of a continuous-time birth-death process, as discussed later in the section.\\

\noindent We now compute the distribution of times to reach the threshold size for different values of the parameter $a$. Figure 7 shows the resulting plots. One finds from the figure that as criticality is approached ($a\to0.5$ from above, for a 2-branch model), the peak of the distribution shifts to a higher value of $t$. This can be understood from the fact that when $a\to0.5$, the competition between production of P and D cells increases, so that it takes more number of generations to reach a given threshold size of P cells. Another notable feature of the approach to criticality is that of a rising variance in the distribution of the extinction time and the time to reach $N_{max}$. The variance as a function of the parameter $\frac{c}{a}$ is plotted in figure 8. The rising variance has been proposed as a signature of regime shift in the dynamics of nonequilibrium systems \cite{scheffer2009early,scheffer2012anticipating,pal2013early}.\\
\begin{figure}
\begin{center}
\begin{minipage}[c]{0.99\linewidth}
\includegraphics[width=13.2cm]{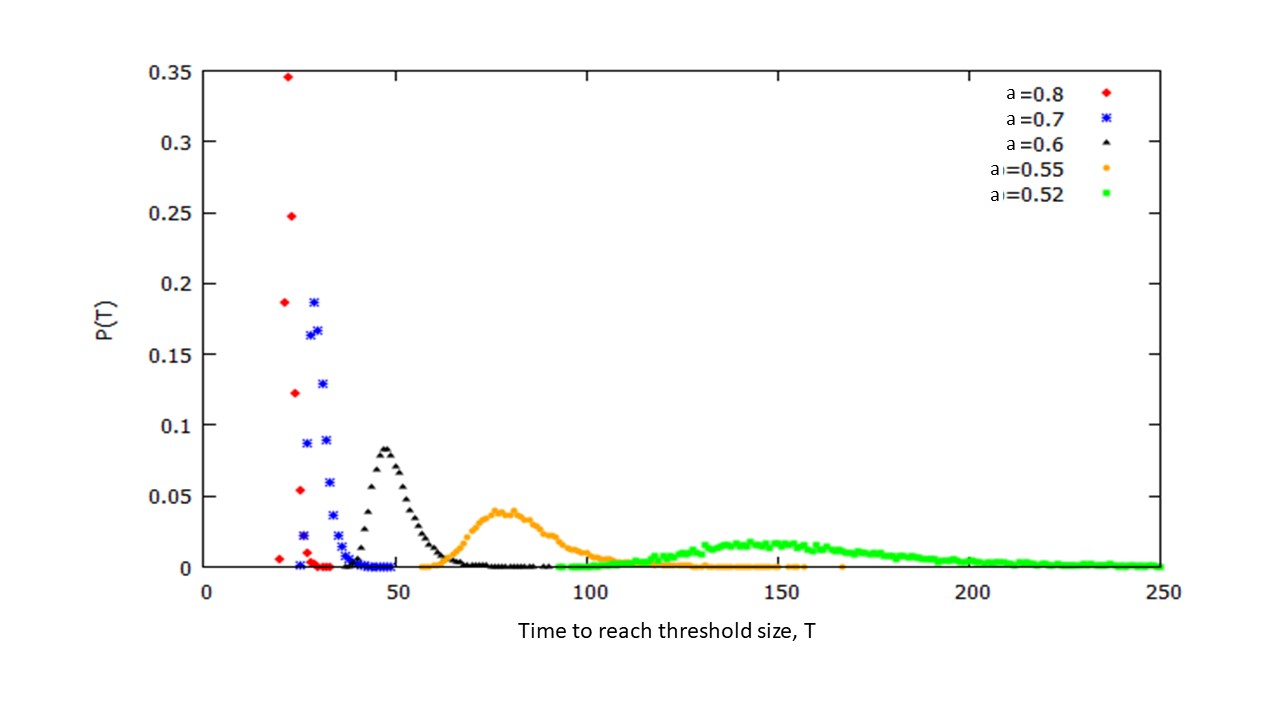}
\end{minipage}
\caption{Simulation results for distribution of times to reach the threshold size for different values of $a>0.5$ and $b=0$ (supercritical branching process). As criticality $(a=0.5)$ is approached, the system takes more and more time to reach the threshold size.}  
\end{center}
\label{DistTthSim}
\end{figure}
\begin{figure}
\begin{center}
\begin{minipage}[c]{0.99\linewidth}
\includegraphics[width=13.2cm]{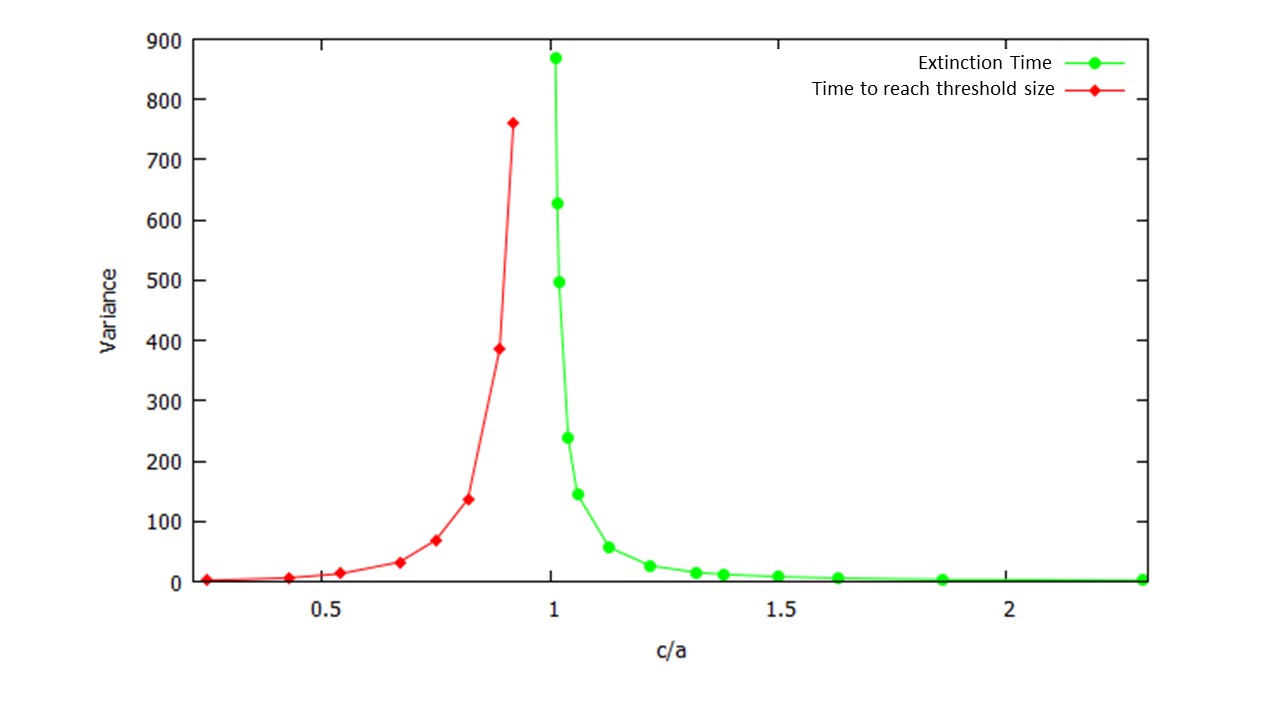}
\end{minipage}
\caption{Plots of variance in the distributions of the extinction time and the time to reach the threshold size versus the parameter $\frac{c}{a}$ $(b=0)$. The variance appears to diverge as the critical point is approached.}  
\end{center}
\label{Variance}
\end{figure}

\noindent So far, we have been considering a DT branching process of the GW type in which time changes in discrete steps. For the case $b=0$, analytic expressions for various quantities can be obtained using the formalism of CT homogeneous Markov process \cite{harris2002theory,zbMATH03410334,kimmelbranching,bailey1990elements,durrett2015branching,karlin1975first}. Let us consider a population of P cells, the total number of which at time $t$ is given by $n(t)$ where time $t$ is now a continuous variable. Each individual in the population is capable of giving birth to new individuals. At the time of birth, a parent may give rise to two offsprings and cease to exist, as in the case of the generation of two P cells through cell division, or the parent may continue to exist along with the offspring as in the case of animal reproduction. Both the descriptions are equivalent in the sense that in each case the total population size of reproducing individuals increases by one. For the CT case, it is more convenient to adopt the second interpretation. In the case of the DT two-branch model, on cell division, the total number of P cells increases by one with probability $a$ (a birth process) and decreases by one with probability $c$ (a death process). In the CT case of the linear birth-death process, let $\lambda\Delta t$ be the probability that an individual gives birth in the time interval $\Delta t$ and $\mu\Delta t$ the probability that the individual dies in time interval $\Delta t$. In the CT branching process theory, $Z(t)$ again represents the population size of the reproducing individuals at time $t$ and the corresponding PGF is defined as
\begin{equation}
    F(s,t)=\sum_{k\geq0}P[Z(t)=k|Z(0)=1]s^k
    \label{eqn21}
\end{equation}
\noindent with $F(s,0)=s$. The PGF of the family size in the case of the two-branch model is
\begin{equation}
    f(s)=\frac{\mu}{\lambda+\mu}+\frac{\lambda}{\mu+\lambda}s^2
    \label{eqn20}
\end{equation}
\noindent Drawing analogies with the discrete-time case (equation (\ref{pgf})), $a=\frac{\lambda}{\mu+\lambda},b=0,c=\frac{\mu}{\lambda+\mu}$. The PGF $F(s,t)$ satisfies the backward Chapman-Kolmogorov (CK) equation \cite{harris2002theory,zbMATH03410334,durrett2015branching}
\begin{equation}
    \frac{\partial F(s,t)}{\partial t}=\mu-(\mu+\lambda)F(s,t)+\lambda F^2(s,t)
    \label{PDF}
\end{equation}
With the initial condition $F(s,0)=s$, the analytic solution of equation (\ref{PDF}) for $\mu \neq \lambda$ is given by
\begin{equation}
    F(s,t)=\frac{\mu(s-1)-e^{-(\lambda-\mu)t}(\lambda s-\mu)}{\lambda(s-1)-e^{-(\lambda-\mu)t}(\lambda s-\mu)}
    \label{eqn24}
\end{equation}
From the PGF, one can obtain the expressions for the probability distributions $p_n (t)=P[Z(t)=n]$ as
\begin{eqnarray}
    p_n(t)=(1-\alpha)(1-\beta)\beta^{n-1},\hspace{20 pt} n\geq1, \nonumber\\
    p_0(t)=\alpha
    \label{CTdiffeqnsol}
\end{eqnarray}
where $\alpha=\frac{\mu(e^{(\lambda-\mu)t}-1)}{\lambda e^{(\lambda-\mu)t}-\mu}$ and $\beta=\frac{\lambda(e^{(\lambda-\mu)t}-1)}{\lambda e^{(\lambda-\mu)t}-\mu}$. From equation (\ref{CTdiffeqnsol}), $p_0 (t)$ yields the probability that the extinction of the population occurs by time $t$ so that the CDF $F_T (t)$, giving the probability that the extinction time $T$ is less than $t$ is $p_0 (t)$. The PDF is obtained by differentiating the CDF with respect to $t$ and the expression for the mean extinction time $T_{ME}$ is given by 
\begin{equation}
    T_{ME}=\int_{0}^{\infty} t\frac{dp_0}{dt} dt=\int_{0}^{\infty} t\frac{d\alpha}{dt} dt 
\end{equation}
In the subcritical case $(\lambda<\mu)$, in which population extinction occurs with probability 1, an analytic expression for $T_{ME}$ can be obtained as
\begin{equation}
    T_{ME}=\frac{1}{\lambda}ln(\frac{\mu}{\mu-\lambda})
    \label{MeanExtTimeTheory}
\end{equation}
Figure 9 shows a plot of $T_{ME}$ versus $\lambda$ which diverges at the critical point $\lambda=0.5$ in contrast with the finite-size effect exhibited in figure 4(a).\\
\begin{figure}
\begin{center}
\begin{minipage}[c]{0.99\linewidth}
\includegraphics[width=13.2cm]{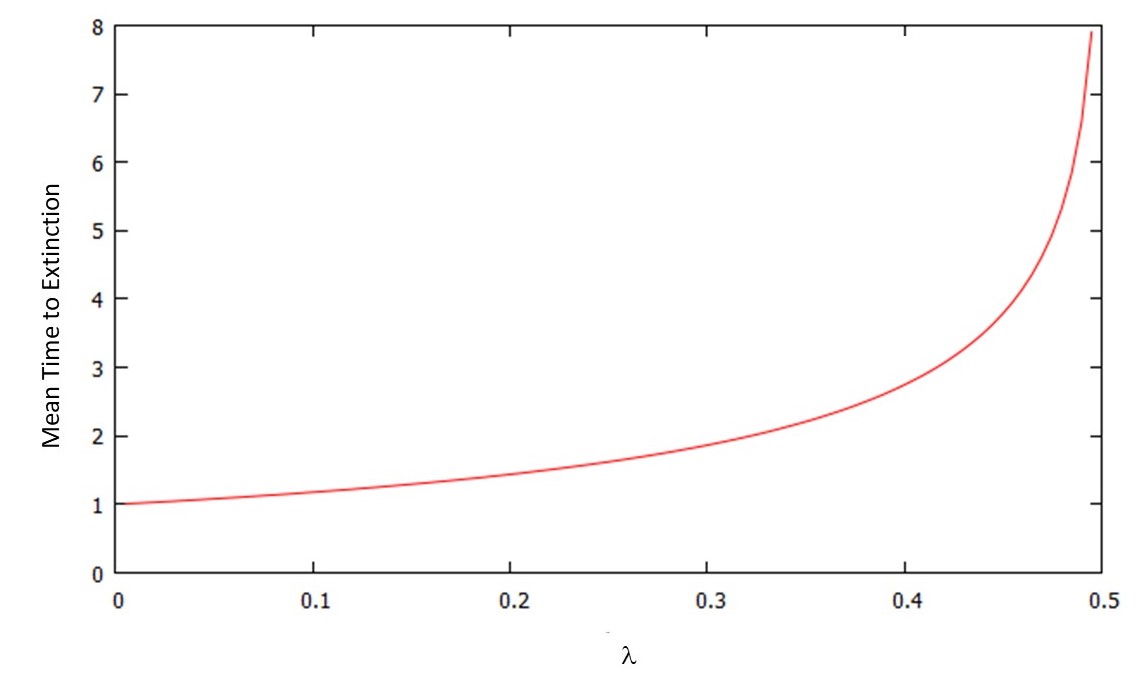}
\end{minipage}
\caption{Plot of the analytic expression of mean extinction time, $T_{ME}$ (equation (\ref{MeanExtTimeTheory})) versus the parameter $\lambda$ $(\lambda<\mu)$. It is seen that $T_{ME}$ diverges as $\lambda\to0.5$.}  
\end{center}
\label{MeanTExtTh}
\end{figure}

\noindent The time to reach the threshold size, $N_{max}$ (conditioned on non-extinction), has a double-exponential (Gumbel) distribution given by
\begin{equation}
    f(t)=\frac{\theta^2N_{max}}{\lambda}\exp(-\frac{\theta}{\lambda}N_{max}e^{-\theta t})e^{-\theta t}
    \label{Gumbel}
\end{equation}
\noindent where $\theta=\lambda-\mu$. This is plotted in figure 10(a) for different values of $\lambda$. The qualitative behaviour matches with that of figure 7 for the discrete time case. The mean time to reach the threshold, computed from equation (\ref{Gumbel}) is
\begin{equation}
    T_{th}=\frac{1}{\theta}log(\frac{N_{max}\theta}{\lambda})+\frac{\gamma}{\theta}
    \label{TheoThMean}
\end{equation}
where $\gamma=0.5772156649$ is the Euler's constant. Figure 10 (b) shows that $T_{th}$ diverges as $\lambda\to0.5$.\\

\begin{figure}
\begin{center}
\begin{minipage}[c]{0.99\linewidth}
\includegraphics[width=13.2cm]{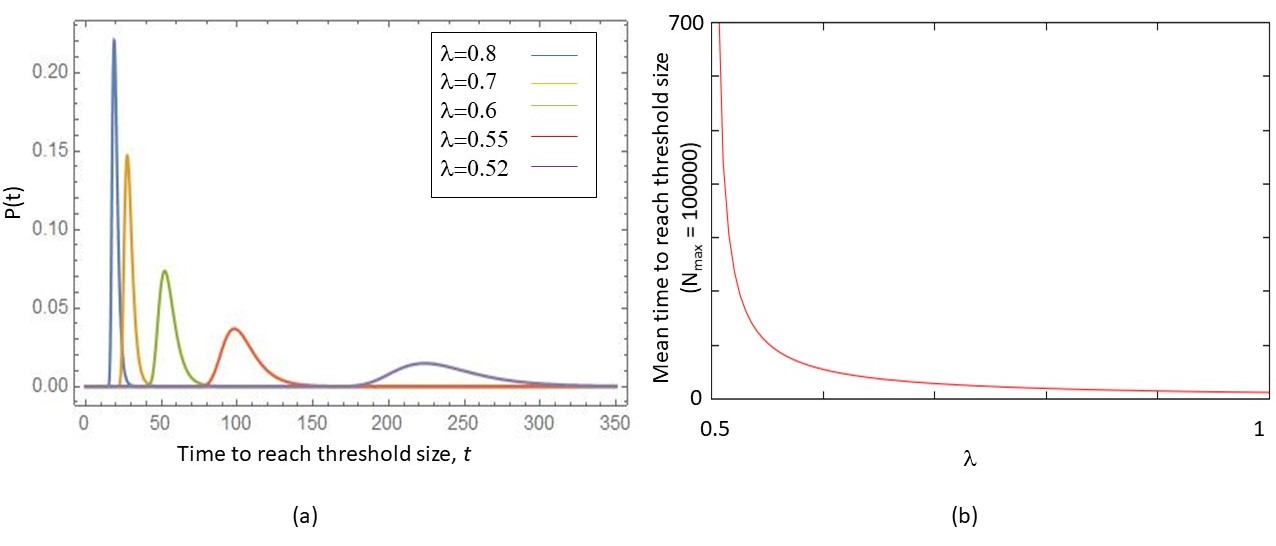}
\end{minipage}
\caption{Plots of: (a) Gumbel distribution of time to reach the threshold size, $N_{max}=100000$ for different values of the parameter $\lambda$ and (b) theoretical mean time to reach the threshold size (equation (\ref{TheoThMean})).}  
\end{center}
\label{DistTthTheo}
\end{figure}
\noindent In the CT case, one can derive a number of analytic expressions for various quantities which is not possible in the case of the DT branching process. In the limit of time $t\to\infty$, the behaviour of $Z(t)$ and associated quantities are very similar in both the cases \cite{zbMATH03410334}. We illustrate this equivalence for two quantities. In the supercritical branching case, the probability of extinction in the large time limit of a CT process is, from equation (\ref{CTdiffeqnsol}), $\frac{\mu}{\lambda}$  which is the same as the expression $\frac{c}{a}$ (equation (\ref{eqn8})) in the DT case. Considering a critical branching process $(\lambda=\mu)$ in the CT case, the probability distributions have the form \cite{harris2002theory,bailey1990elements}
\begin{eqnarray}
    p_n(t)=\frac{(\lambda t)^{n-1}}{(1+\lambda t)^{n+1}}, \hspace{20 pt} n\geq1,\nonumber\\
    p_0(t)=\frac{\lambda t}{(1+\lambda t)}
    \label{CTdiffeqnsolcrit}
\end{eqnarray}
\noindent In the asymptotic limit of $t\to\infty$, the PDF of extinction times is
\begin{equation}
    p_e(t)=\frac{dp_0}{dt}\sim \frac{1}{t^2}
    \label{DistTex}
\end{equation}
\noindent We next consider the DT branching process. The exact time of extinction $T=t$ if the size of the population becomes zero for the first time in generation $t$. This implies the conditional statement $Z_t=0 \hspace{4 pt}\cap\hspace{4 pt} Z_{t-1}>0$. One can further write
\begin{equation}
    P(Z_t=0 \hspace{4 pt}\cap\hspace{4 pt} Z_{t-1}>0) + P(Z_t=0 \hspace{4 pt}\cap\hspace{4 pt} Z_{t-1}=0) = P(Z_t=0)
\end{equation}
\noindent The second term on the l. h. s. can be written as $P(Z_{t-1}=0)$ since $Z_t$ is necessarily zero if $Z_{t-1}=0$. Thus, the distribution of the extinction time $T$ is given by
\begin{equation}
    P_e(T=t) = P(Z_t=0) - P(Z_{t-1}=0)
\end{equation}
\noindent From Theorem 1 (equation (\ref{Th1})) for the critical branching process, $P(Z_t>0)=\frac{2}{t}$ in the large time limit with $a=0.5$ in the two-branch model. Thus, in the large time limit,
\begin{equation}
    P_e(T=t)=(1-\frac{2}{t})-(1-\frac{2}{t-1})\sim \frac{2}{t^2}
    \label{ExTDistcrit}
\end{equation}
\noindent in agreement with the result (equation (\ref{DistTex})) for the CT branching process. We will revisit the last result in the next section. We also point out that the experimental results on tissue homeostasis have been explained earlier by making use of the CT probability distributions shown in equation (\ref{CTdiffeqnsolcrit}) \cite{klein2011universal,clayton2007single,klein2007kinetics}. In the limit $t\to\infty$, it is straightforward to verify that the DT branching process results, as contained in Theorems 1, 2 and 3, reproduce those obtained in the CT case to describe the experimental results.\\

\noindent We end this section by showing that a quantitative comparison of the simulation results in the case of the two-branch model describing a DT GW process with analogous analytic expressions in the CT case is possible due to the property of embeddability. In Appendix A, the embeddability criterion is discussed with the demonstration that the DT GW process, described by the two-branch model, is embeddable in the CT linear birth-death process. From equation (\ref{eqn44}) of Appendix A, one finds that
\begin{equation}
    \frac{a}{c}=\frac{\lambda}{\mu}, \hspace{5 pt} c=1-a
    \label{eqn46}
\end{equation}
In figure 11(a), the simulation data are fitted with the analytic expression for the mean extinction time $T_{ME}$ (equation (\ref{MeanExtTimeTheory})) in the subcritical case $\lambda<\mu$, i.e., $a<0.5$. The argument of the logarithm in equation (\ref{MeanExtTimeTheory}) is a function of $a$ using the relations in equation (\ref{eqn46}). The pre-factor $\frac{1}{\lambda}$ of the logarithm is not a unique function of $a$ (only the ratios $\frac{a}{c}$ and $\frac{\lambda}{\mu}$ are fixed) so that $\lambda$ can be treated as a free parameter. The best fit between the simulation data and the analytic expression is obtained for $\lambda=\frac{a}{0.91}$. Figure 11(b) compares the simulation data of figure 6(a) with the analytic expression for the mean time $T_{th}$ to reach the threshold population size (equation (\ref{TheoThMean})). The analytic formula is re-expressed in terms of the parameter $c$ treating $\mu$ as a free parameter. The expression for $T_{th}$ is given by
\begin{equation}
    T_{th}=\frac{1}{\mu}\frac{c}{1-2c} log (\frac{N_{max}(1-2c)}{1-c}) + \frac{\gamma}{\mu}\frac{c}{1-2c}
\end{equation}
The best fit is obtained for $\mu=\frac{c}{1.1}$. Figure 11(c) shows the simulation data for $P(t)$, the distribution of times to reach the threshold size, fitted by the analytic expression in equation (\ref{Gumbel}). The parameter $\mu$ was eliminated using the constraints imposed by equation (\ref{eqn44}), in favour of $\lambda$ and $a$. For a given value of the parameter $a$, $\lambda$ was used as the fitting parameter.
\begin{figure}
\begin{center}
\begin{minipage}[c]{0.99\linewidth}
\includegraphics[width=13.2cm]{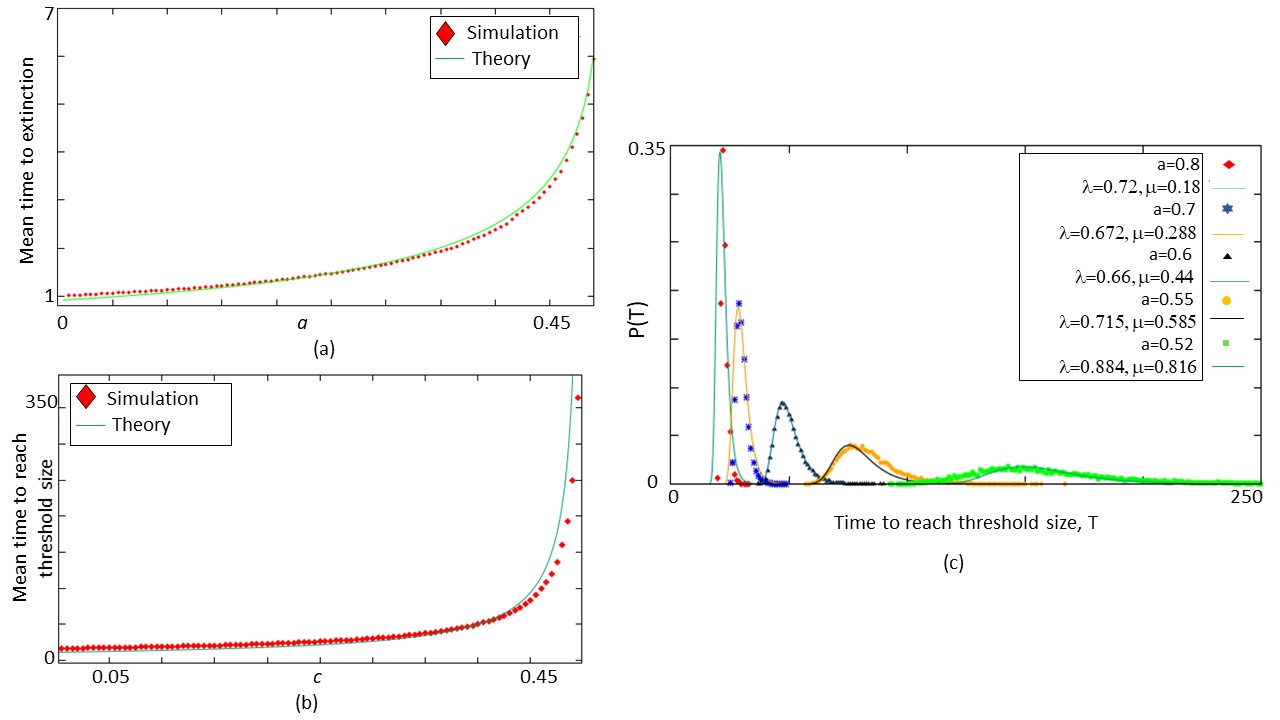}
\end{minipage}
\caption{Comparison of analytic results for CT birth-death process and simulation results for DT GW process. (a) Mean extinction time in the subcritical case ($a<0.5$); (b) Mean time to reach the threshold size ($N_{max}=100,000$); (c) Distribution of times to reach the threshold size. In all the cases, continuous solid lines represent theoretical results whereas points represent simulation outcomes.}    
\end{center}
\label{Comparison}
\end{figure}

\section{Branching Process, Avalanche and Percolation Model}
\noindent The problem of tissue homeostasis finds a natural representation in a branching process \cite{roshan2014exact,rue2015cell} with homeostasis signifying criticality. Several studies have established a correspondence between a critical branching process and criticality in sandpile and percolation models in the mean-field limit \cite{alstrom1988mean,zapperi1995self,gros2010complex,garcia1993branching,lee2004branching}. We exploit these analogies to describe tissue homeostasis in terms of critical quantities characterising the lattice-statistical models.\\

\noindent In the percolation model, a disordered system is described as a network of elements (sites or bonds). The probability that a site (site percolation) or a bond (bond percolation) is present is $p$. When $p=0$, the network does not exist. For small values of $p$, the network is fragmented, whereas the network is fully connected when $p=1$. A critical point transition occurs at the percolation threshold $p_c (0<p_c<1)$ such that for $p>p_c$, a long-range connectivity is established across the system. Below $p_c$, finite-sized clusters of connected elements coexist whereas above $p_c$, a giant cluster (infinite cluster) spanning the system coexists with smaller-sized clusters. The critical point transition is characterised by critical phenomena occurring at or close to the critical point \cite{sornette2006critical,christensen2005complexity}. The most prominent feature among these is the appearance of power-law singularities in cluster-related quantities close to $p_c$. For example, the average cluster size diverges as  $S_{av}\sim|p-p_c|^{-\gamma}$ in the critical region with $\gamma$ defining a critical exponent.\\

\noindent The phenomenon of self-organised criticality (SOC) is wide-spread in nature with the sandpile model serving as a well-known paradigm \cite{sornette2006critical,dhar1990abelian}. In the sandpile model defined on a lattice, the pile is generated through additions of sand particles at random sites. If the height of the pile at a site reaches a critical value, a toppling occurs at the site transferring sand particles to each of the neighbouring sites. This continues in successive time steps till all the sites have sand piles with height less than the critical height. The sequence of topplings constitutes an avalanche. The size of an avalanche is given by the number of sites which topples during the lifetime of the avalanche with the size-distribution $D(s)$ obeying a power-law, $D(s)\sim s^{-\tau}$, in the self-organised critical state. One can also define the duration of the avalanche defined by the number of time steps through which the avalanche progresses before coming to a stop. The avalanche duration, $D(T)$, also has a power-law form, $D(T)\sim T^{-\delta}$ in the self-organised critical state.\\

\noindent The mean-field theory (MFT) of lattice statistical models exhibiting critical point transitions is equivalent to studying the models on the Bethe lattice which has a branching structure and effective dimension $d\to\infty$. It has been shown earlier that the Abelian sandpile model (the order of topplings is immaterial) of SOC on the Bethe lattice has critical exponents which are the same as those of the mean-field percolation model \cite{dhar1990abelian}. We now draw on the analogies between a branching process and avalanche and percolation models on the Bethe lattice to point out that all the three models exhibit similar critical behaviour. Towards this goal, we first derive the PGF of the total progeny distribution in a branching process. We define a random variable $X$ which counts all the P cells including the founding cell.  Thus, in a DT branching process,
\begin{equation}
    X=\sum_{t\geq0}Z_t=1+\sum_{t\geq1}Z_t
\end{equation}
\noindent In figure 2, the total number of P cells is ten up to $t = 4$. The PGF of $X$ is defined as
\begin{equation}
    g(s)=\sum_{k\geq1}P(X=k)s^k
    \label{pgfdef}
\end{equation}
\noindent The PGF $G(s)$ of the offspring distribution (family size) is as given in equation (\ref{pgf}). One can show that $g(s)$ is given by the solution of the equation \cite{feller2008introduction}
\begin{equation}
    g(s)=sG(g(s))
\end{equation}
With the form of $G(s)$ known, one can derive an expression for $g(s)$ as
\begin{equation}
    g(s)=\frac{1-bs-\sqrt{(bs-1)^2-4acs^2}}{2as}
    \label{pgf2}
\end{equation}
\noindent The criticality condition for the three-branch model is given by $a=c=a_c$. The survival probability $P_s$ of the population of P cells serves as an order parameter of the critical point transition with 
\begin{eqnarray}
    P_s\neq0,\hspace{10 pt} a>a_c \nonumber \\
    P_s=0, \hspace{10 pt} a\leq a_c
\end{eqnarray}
\noindent For $a>c$, $P_s$ is given by $1-\frac{c}{a}$. Close to the critical point, $P_s$ has the power-law form
\begin{equation}
    P_s\sim\frac{1}{a_c}(a-a_c)^\beta, \hspace{5 pt} \beta=1
\end{equation}
\noindent Let $\langle k \rangle$ be the average size of the clusters of P cells. In the subcritical regime, the average size diverges as 
\begin{equation}
    \langle k \rangle=\frac{\partial g}{\partial s}|_{s=1}\sim (a_c-a)^{-\gamma}
\end{equation}
as $a\to a_c$ with $\gamma=1$. In the supercritical regime, considering only extinct cell populations, the average cluster size has the power-law form
\begin{equation}
    \langle k\rangle \sim (a-a_c)^{-\gamma^*}
\end{equation}
as the critical point is approached with $\gamma^*=1$.  For the two-branch model $(b=0)$,  one has $a=p, c=1-p$ with the critical point defined by $a_c=p_c=\frac{1}{2}$. One can easily check that the critical exponents have the same values as in the case of the three-branch model, indicating universality of critical phenomena. The exponents $\beta, \gamma, \gamma^* (\gamma=\gamma^*)$ have values identical to the exponents associated with the order parameter and the average cluster size respectively in the mean-field percolation model.\\

\noindent In the branching process depicting the proliferation of P cells, the spreading of the proliferation activity through subsequent generations is analogous to the spreading of an avalanche in the sandpile model. The correspondence can be clearly understood by considering the two-branch model of P cell proliferation. In each generation, a P cell is replaced by two P cells with probability $p$ and it does not leave P cells as descendants with probability $1-p$. In terms of an avalanche, an active site relaxes (``topples") with probability $p$ giving rise to two new active sites. The probability that the active site does not relax, i.e., no further active site is generated is $1-p$. The process is repeated for each new active site resulting in the spreading of the avalanche. The avalanche comes to a stop when the number of new active sites falls to zero. The regime $p<p_c$ corresponds to the subcritical regime in the branching process (population extinction occurs with probability one) corresponding to solely finite-sized avalanches in the sandpile model and finite-sized clusters in the percolation model. On the other hand, in the supercritical region $(p>p_c)$, the probability of having an infinite population/avalanche/cluster size is non-zero. The PGF for the two-branch model is obtained by putting $b=0$ in the expression for $g(s)$ in equation (\ref{pgf2}). By expanding the PGF in powers of $s$ and comparing with the expression in equation (\ref{pgfdef}), one obtains the following results as $p\to p_c=\frac{1}{2}$ from below:
\begin{eqnarray}
    P(k,p)\sim k^{-\tau_k} \exp(-\frac{k}{k_c})\\
    k_c(p)\sim |p-p_c|^{-\frac{1}{\sigma}}
\end{eqnarray}
with $\tau_k=\frac{3}{2}$ and $\sigma=\frac{1}{2}$. Also, the size distribution at the critical point is given by 
\begin{equation}
    P(k,p_c) \sim k^{-\tau_k}
    \label{eqn58}
\end{equation}
\noindent The distribution captures the power-law form of the avalanche size distribution in the self-organized critical state with the value $\tau_k=\frac{3}{2}$ the same as the mean-field estimate. Furthermore, the extinction time distribution (equation (\ref{ExTDistcrit})) reproduces the avalanche lifetime distribution $D(T)\sim T^{-\delta}, \delta=2$ in MFT. The value of $\sigma=\frac{1}{2}$ also agrees with the mean-field estimates. The results can be generalised to the three-branch model with identical values of the critical exponents. In the experiments on tissue homeostasis \cite{clayton2007single,klein2007kinetics}, the lineage tracing technique keeps track of the progeny of labelled cells. The technique, however, is unable to distinguish between the P and D cells so that the total count of cells includes both the P and D cells. This, however, does not pose a problem when experimental observations are compared with branching process results as shown below. In the latter case, the population consists of solely P cells.\\

\noindent The total number of cells (P+D) in the $t$-th generation is $C_t=2Z_{t-1}$ where $Z_{t-1}$ is the number of P cells in the $(t-1)$-th generation (the D cells do not reproduce). From equation (\ref{cdf}), one can obtain the probability distribution of P cells at large time $t-1$ (conditioned on non-extinction) as
\begin{eqnarray}
    P(Z_{t-1}=u)=P(Z_{t-1}\leq u)-P(Z_{t-1}\leq u-1) \nonumber \\
    \hspace{55 pt} \sim \frac{1}{a(t-1)}\exp (-\frac{u}{a(t-1)})
\end{eqnarray}
which leads to ($t$ is large)
\begin{equation}
    P(C_t=2u)\approx \frac{1}{at}\exp(-\frac{2u}{2at})
    \label{eqn47}
\end{equation}
Equation (\ref{eqn47}) shows that the size distribution of the total number of cells has the same scaling form as in the case of P cells.\\

\noindent For both the two-branch and three-branch models, a simple counting argument \cite{christensen2005complexity} shows that the size (number of cells) of the D cell population is equal to $k+1$ where $k$ is the size of an extinct population of P cells. The total number of cells is thus $2k+1$. Thus for large $k$, the size distribution of the total number of cells (the experimentally measurable quantity) at the critical point has the same power-law form, as shown in equation (\ref{eqn58}), with the same magnitude of the critical exponent. The simple relationship gives rise to the possibility of testing the power-law forms of the size and lifetime distributions of the descendant cells in lineage tracing experiments.\\
\section{Concluding Remarks}
\noindent The maintenance of adult tissues in the homeostatic condition is an essential requirement for the structural and functional integrity of an organism. In the adult stage, deviations from the condition occur due to external injuries or due to an abnormal proliferation of cells as in the case of cancer. In the first case, wound healing processes set in to restore the homeostatic condition whereas in the second case, therapeutic interventions are needed to restore the balance. The problem of tissue homeostasis involving an exquisite balance between cell proliferation and cell loss offers an ideal opportunity for applying the concepts and techniques of nonequilibrium statistical physics to investigate how the crucial balance is achieved. We have utilised the theorems and techniques of  branching process theory to show that the basic experimental observations on the homeostasis of mouse epidermis \cite{klein2011universal,jones2008opinionepidermal,clayton2007single,klein2007kinetics} can be understood in terms of a critical branching process. The critical state is at the border between the subcritical and supercritical regions with the probability of extinction of the population of P cells serving as an order parameter. Through numerical simulation as well as analytic results we have obtained a number of temporal signatures of the approach to criticality which could be tested in appropriately designed experiments. A quantitative comparison between simulation results in the case of the two branch model with the analytic expressions obtained in the CT case could be carried out due to the special feature of embeddability. This is one of the exceptional cases in which a comparison of discrete and continuous-time results can be  meaningfully compared. We have further drawn on the equivalence between the critical branching process and the mean-field avalanche and percolation models to show that the size and lifetime distributions of the population of P cells approaching the critical point have power-law forms. The associated critical exponents have magnitudes equal to the mean-field estimates. The value of the size distribution exponent $\tau_k=\frac{3}{2}$ (equation (\ref{eqn58})) is also stipulated by a branching process theorem \cite{harris2002theory}. Lineage tracing experiments on tissue homeostasis could be designed to test the power-law predictions. The two- and three-branch models of tissue homeostasis exhibit the same critical behaviour signifying universality, a key feature of critical phenomena. In the critical state, the probability distribution of the population size attains an invariant scaling form in the long-time limit consistent with experimental observations \cite{clayton2007single,klein2007kinetics}.\\

\noindent In most of the lineage tracing experiments carried out so far, fixed samples were taken at different time points so that an individual progenitor cell could not be tracked over time. Rompolas et al. \cite{rompolas2016spatiotemporal} used two-photon microscopy in conjunction with live imaging to follow individual cells through their lifetimes enabling them to offer new insights on epidermal homeostasis. In contrast to the earlier studies in which asymmetric division $(P\rightarrow PD)$ was found to be the predominant mode of cell division, the study using live imaging in the ear and paw of mice epidermis showed that there was an almost 50:50 chance of every cell undergoing direct differentiation or undergoing cell division to produce two P cells. The experiment revealed that the cell behaviour is not coordinated between generations and sibling lifetimes are coupled. The findings add relevance to the two-branch model of tissue homeostasis. The critical behaviour of the branching process models is that of the birth-death process. In these models, the proliferation and differentiation kinetics are intracellular (cell-autonomous). The models have been designated as zero-dimensional to indicate that the spatial distribution of cells and cell-cell interactions are not taken into account. Some studies on tissue homeostasis put focus on intercellular interactions as the key driver of cell fate decisions \cite{klein2011universal,mesa2018homeostatic,yamaguchi2017dynamical,klein2008mechanism}. The key assumption in a spatial model of cellular kinetics is that the P cells divide only when a neighbouring differentiated cell migrates to the suprabasal layers \cite{klein2008mechanism}. This conjecture is supported by recent experimental evidence \cite{mesa2018homeostatic}. In the case of cell-intrinsic regulation described by a critical birth-death  process, the average size of the surviving clones grows as $\langle n(t)\rangle \sim t$ and the clone size acquires a scaling form described by the scaling function $F(x)=\exp(-x)$ in the large time limit. In the case of cell extrinsic regulation in which spatial considerations are important, the scaling forms are $\langle n(t)\rangle  \sim \sqrt{t}$, $F(x)\sim e^{-\frac{\pi x^2}{4}}$ in one dimension (1d). The results are consistent with experimental measurements in 1d tissues like intestinal crypts \cite{lopez2010intestinal} and seminiferous tubules \cite{klein2010mouse}. In 2d, $\langle n(t)\rangle \sim t$  with logarithmic corrections and $F(x)=\exp(-x)$. In dimension $d\geq3$, the scaling forms are the same as in the case of cell intrinsic regulation. The scaling forms in the case of the cell extrinsic regulation are derived from the voter model (VM) in which the opinion of an agent is influenced by that of a neighbour \cite{klein2011universal,yamaguchi2017dynamical,klein2008mechanism}. The characteristic features of the clonal dynamics in experimental investigations of skin tissues (2d systems) are reproduced well by both cell intrinsic and cell extrinsic regulation models. Critical phenomena in living systems constitute a newly emerging research with an interdisciplinary character \cite{mora2011biological,munoz2018colloquium,pal2014non,bose2017criticality,bose2019bifurcation}. The emergence of universal features in living systems close to criticality is captured by statistical physics models, which elucidate the basic principles governing the critical behaviour of a large class of systems.\\  
\appendix
\section
\noindent A CTM branching process is a sequence of transitions or jumps between states separated by random time intervals known as waiting or sojourn times which are exponentially distributed. In the DT branching process of the GW type the jumps occur at fixed intervals of time. Every CTM process has a DT process embedded in it if only the jump events are considered, ignoring the randomly distributed waiting times between the jumps. The discrete process is of the GW type if the time intervals between successive jumps are fixed to be $\delta$. The converse question of whether a DT branching process with a specific offspring PGF $f(s)=\sum_{j=0}^{\infty} p_{j}s^j$ is embeddable in a CTM process  is more problematic \cite{harris2002theory,zbMATH03410334,karlin1975first}. The embeddability criterion stipulates that a PGF $f(s)$ is embeddable if there exists a PGF $F(s,t)$, defined in equation (\ref{eqn21}), such that $F(s,t+u)=F(F(s,t),u); t,u \geq0,|s|\leq1$ and $F(s,\delta)=f(s)$ for some $\delta>0$. Using this criterion, most of the familiar PGFs turn out to be nonembeddable. The linear fractional GF $g(s)$ is an exception and has the form
\begin{equation}
    g(s)=\frac{\alpha+\beta s}{\gamma+\sigma s}, \hspace{10 pt} \alpha\sigma - \beta\gamma \neq 0
\end{equation}
It is easy to check that the successive iterates of $g(s)$ have the linear fractional form (LFF). In the case of the CT linear birth-death process, the PGF given by equation (\ref{eqn24}) is of the LFF such that $F(s,t+u)=F(F(s,t),u)$ has the same form as $F(s,t)$. Writing $t$ as $t=n\delta, n=0,1,2,...$, the $n^{th}$ iterate of $F(s,\delta)$ yields $F(s,t)=F(s,n\delta)$. We will now show that for the linear birth-death process and for infinitesimal $\delta$, $F(s,\delta)=f(s)$, the offspring PGF of the two-branch model given by equation (\ref{pgf}) with $b=0$. The embeddability makes it possible to compare the simulation results of the DT case with the analytic results obtained in the case of the CT process.\\

\noindent A natural description of a CTM process is provided by the infinitesimal GF $u(s)=\sum_{k=0}^{\infty}d_k s^k$.The infinitesimal probabilities of the process are represented by the expression $\delta_{1k}+d_k h+o(h)$, where $\delta_{1k}$ is the Kronecker delta symbol. The coefficients $d_k$'s satisfy the relations $d_1\leq0, d_k\geq0 (k=0,2,3,..)$ and $\sum_{k=0}^{\infty}d_k =0$. For $d_k\geq0$, $d_k h$ specifies the probability that a single individual is replaced by $k$ individuals in the time interval $(t,t+h)$. The Markov process is assumed to be temporally homogeneous so that the coefficients $d_k$'s do not depend on time. In terms of the infinitesimal probabilities, the physical characterization of the CT process is as follows. The lifetime of an individual is a random variable with exponential distribution. The mean lifetime is given by
\begin{equation}
    \lambda_m^{-1}=d_0 + d_2 + d_3 + ...
\end{equation}
At the end of its lifetime, an individual produces a random number $X$ of offspring described by the probability distribution
\begin{equation}
    Pr\{X=k\}=\frac{d_k}{d_0+d_2+d_3+...}, \hspace{10 pt} k=0,2,3,...
\end{equation}
In the case of the linear birth-death process, one has $d_2=\lambda$ , $d_0=\mu$, $d_1=-(\lambda+\mu)$ and $d_k=0$ otherwise. Also, $\frac{\lambda}{\lambda+\mu} (\frac{\mu}{\lambda+\mu})$ is the probability of a birth(death) at the occurrence of an event.\\

\noindent The PGF $F(s,t)$ (equation (\ref{eqn21})) can be rewritten as
\begin{equation}
    F(s,t)=\sum_{k=0}^{\infty} P_{1k}(t)s^k
\end{equation}
where $P_{1k}$ represents the transition probability from state 1 (one individual) to state $k$ ($k$ individuals). For an infinitesimal time interval $h$, one can write
\begin{eqnarray}
 F(s,h)=\sum_{k=0}^{\infty} P_{1k}(h)s^k= \sum_{k=0}^{\infty}(\delta_{1k}+d_k h + o(h))s^k \nonumber\\
 \hspace{30 pt} = s+hu(s)+o(h)
 \label{eqn39}
\end{eqnarray}
In the case of the linear birth-death process, putting $h=\delta$ in the expression for $F(s,t)$ (equation (\ref{eqn24})) and ignoring terms of the order of $o(\delta)$, one gets from equation (\ref{eqn39}) the following expression for the infinitesimal GF $u(s)$:
\begin{equation}
    u(s)=\lambda s^2 - (\lambda+\mu)s + \mu
\end{equation}
The infinitesimal GF can further be rewritten as
\begin{equation}
    u(s)=a(f(s)-s)
\end{equation}
where
\begin{equation}
    a=\lambda + \mu
\end{equation}
and $f(s)$ is the PGF given by equation (\ref{eqn20}). The PGF has the same form as that of the offspring PGF of the DT two-branch model given by equation (\ref{pgf}) with $b=0$. As pointed out earlier, the correspondence between the two parameter sets is given by
\begin{equation}
    a=\frac{\lambda}{\lambda+\mu}, \hspace{10 pt} c=1-a=\frac{\mu}{\lambda+\mu}
    \label{eqn44}
\end{equation}
One can further check from the expression for $F(s,t)$ (equation (\ref{eqn24})) that for small $\delta$,
\begin{equation}
    F(s,\delta)=\lambda s^2 \delta + (s-(\lambda+\mu)s\delta) + \mu \delta + o(\delta)
    \label{eqn45}
\end{equation}
In the case of a DT GW process, the generation time is fixed at the value $\delta$ with $\delta=\frac{1}{\lambda+\mu}$ since birth/death events occur only at the end of a generation. On substituting the value of $\delta$ in equation (\ref{eqn45}), one recovers the expression for $f(s)$ in equation (\ref{eqn20}), i.e., $F(s,\delta)=f(s)$, the embeddability condition discussed earlier.\\

\section*{Acknowledgement}
IB acknowledges the support by CSIR, India, vide sanction Lett.No. 21(0956)/13/EMR-II dated 28.04.2014. The authors thank Sayantari Ghosh for her help in preparing the manuscript. SG would like to thank Parongama Sen for useful discussions.

\section*{References}

\bibliographystyle{iopart-num}
\bibliography{name.bib}

\end{document}